\documentstyle[floats,aps,preprint,prl]{revtex}
\begin{document}
\draft

\title{Quantum disordered phase in a doped antiferromagnet}
\author{C. K\"ubert and A. Muramatsu}
\address{Institut f\"ur theoretische Physik der
Universit\"at W\"urzburg, Am Hubland, \\
97074 W\"urzburg, Germany}
\date{\today}
\maketitle

\begin{abstract}
A quantitative description of the transition to a quantum disordered phase in
a doped antiferromagnet is obtained with a U(1) gauge-theory, where the gap
in the spin-wave spectrum determines the strength of the gauge-fields. They
mediate an attractive long-range interaction whose possible bound-states
correspond to charge-spin separation and pairing.
\end{abstract}
\pacs{PACS numbers: 71.27.+a, 74.20.Mn, 75.10.Jm}

Based on the previous pioneering
analysis of the order-disorder transition in a quantum
antiferromagnet \cite{chn,sach}, a phase diagram for the high-T$_{\rm c}$
cuprates (HTC) was proposed recently
\cite{sopi} that gives a unified view \cite{barz}
of nuclear relaxation, magnetic susceptibility and neutron scattering
experiments, with a spin-gap in the quantum disordered (QD) phase. In spite
of these virtues, the connection with the doped materials remains
phenomenological and moreover, the question of superconductivity remains
unaddressed.
\par
It is shown here, that a) a quantitative description of the
transition to the QD phase can be obtained on the basis of a realistic,
microscopic model of the HTC, b) the mass-gap of the spin-wave excitations in
the QD phase gives the strength of a gauge-field that mediates a long-range
interaction among dopant holes and S-$\frac{1}{2}$ magnetic excitations, and
c) the possible bound states correspond to charge-spin separation and pairing.
Thus, a basis is provided to the phenomenological description of Refs.\
\cite{sopi,barz} and the relationship to charge-spin separation and pairing
is revealed.
\par
The underlying microscopic model is the spin-fermion (SF) one \cite{sfm} that
results from a strong-coupling expansion of the 3-band Hubbard model
\cite{emer,varm}. It is characterized by a Kondo-like coupling of the dopant
holes to localized
spins on the Cu-sites with a strength $J_K$ and a Heisenberg
antiferromagnetic (AF) interaction among the Cu-spins with strength $J_H$.
The long-wavelength limit of the model assuming short-range AF order is given
by \cite{prb}:
\begin{eqnarray}
\label{labor}
S&=&S_F+S_{NL\sigma} \; ,
\\
\label{fermlab}
S_F&=&
\int_0^\beta d\tau \int d^2r \: \biggl[c^{\dagger} \partial_\tau c
+ \frac{1}{2m_r}\;\partial_r c^{\dagger}
\; \partial_r  c
- \gamma_r\; (\partial_r \vec n)^2 \; c^{\dagger} c
+i \sum_\mu \gamma_\mu \; \vec J_\mu^S \cdot \vec J_\mu^F \biggr]
\\
\label{nlsmlab}
S_{NL\sigma}&=&\frac{1}{2g}\int_0^\beta d\tau \int d^2r \:
\biggl[
\left(\partial_r \vec n \right)^2+
\frac{1}{c^2} \left(\partial_\tau\vec n \right)^2
\biggr]
\; ,
\end{eqnarray}
where $c^{\dagger} =(c_\downarrow^\dagger , c_\uparrow^\dagger)$
is a fermionic spinor and $\vec n$ denotes the order parameter of the
spin background, whose dynamics is given by the $O(3)$
non-linear $\sigma$ model
(\ref{nlsmlab}). The first two terms in $S_F$ describe the kinetics of the
fermions, where for low doping the bare band-structure is given by an
anisotropic effective mass $m_r$, with $r=\perp,\parallel$ denoting the
directions perpendicular and parallel to the border of the magnetic
Brillouin-zone around the points $\vec k_{min}=
(\frac{\pi}{2},\pm \frac{\pi}{2})$. Such hole pockets result from the assumed
short-range AF order of the spin-background \cite{prb}. The third term gives
a renormalization of the spin-stiffness of the
non-linear $\sigma$ model, whenever
the corresponding site is occupied by a hole. The fourth term
gives a current-current coupling similar to the one obtained
by Shraiman and Siggia for the $t$-$J$ model \cite{shra1}. The temporal
component ($\mu=\tau$) corresponds to a coupling of the local spin-density
$\vec J_\tau^F=c^\dagger\vec \sigma c$ of the fermions to the background
magnetization $\vec J_\tau^S = \vec n\times \partial_\tau \vec n$, whereas
the spatial components corresponds to the coupling between
the spin current of the holes
$\vec J^F_r= (\partial_r c^\dagger\vec\sigma c -
c^\dagger\vec\sigma\partial_r c)$
and the magnetization current of the background
$\vec J_r^S= \vec n\times \partial_r \vec n$. The connection of the parameters
$m_r$ and $\gamma_\mu$ to the microscopic ones of the SF model and a detailed
derivation of the above results is given in Ref.\ \cite{prb}. For simplicity
we assume here isotropic parameters $m_\perp=m_\parallel=m$ and
$\gamma_\perp=\gamma_\parallel=\gamma$ in the following \cite{note1}.
Further insight is obtained by rotating the fermionic spinor to a
spin quantization axis (SQA), which is given by the local direction of the
order parameter field $\vec n(\vec r,\tau)$:
\begin{eqnarray}
\label{rotation}
p(\vec r,\tau)&=& U^\dagger  (\vec r,\tau) \; c(\vec r ,\tau)\; ,\\
\sigma^z &=&
U^\dagger (\vec r,\tau) \vec \sigma\cdot \vec n (\vec r,\tau)
U (\vec r,\tau)
\; ,
\end{eqnarray}
where $U (\vec r,\tau)$ is an $SU(2)$ rotation matrix.
In the following we use a $CP^1$ representation
\cite{raja} for the rotation matrix
$U$:
\begin{eqnarray}
\label{cp1reprensentation}
U= \left( \begin{array}{cc}
               z_1  &  -\bar z_2      \\
               z_2  &   \bar z_1
             \end{array}
      \right) \; ,
\end{eqnarray}
with $\bar Z Z = 1$, $\bar Z = (\bar z_1,\bar z_2)$,where $z_i$
are complex numbers.
The transition from the action in the laboratory reference frame (\ref{labor})
with a fixed SQA
to a field theory in the rotating reference frame is
achieved by the following set of transformation equations:
\begin{eqnarray}
\label{trafo}
\mbox{rotating SQA} &\longleftrightarrow& \mbox {uniform SQA} \nonumber \\
p_\mu &\longleftrightarrow& c_\mu \nonumber \\
\partial_\mu + i A_\mu \sigma^z &\longleftrightarrow&
\partial_\mu - \frac{i}{2}\vec \sigma \cdot
                 \left( \vec n \times \partial_\mu \vec n \right) \\
K_\mu=i B_\mu \sigma^{-} + i \bar B_\mu \sigma^{+} &\longleftrightarrow&
\frac{i}{2}\vec \sigma \cdot \left( \vec n \times
\partial_\mu \vec n \right) \; ,
\nonumber
\end{eqnarray}
where we have introduced a composite gauge field $A_\mu$ and off-diagonal
contributions denoted by $B_\mu$ and $\bar B_\mu$:
\begin{eqnarray}
\label{gauge-fields}
A_\mu&=&-i\bar Z\partial_\mu Z  \\
B_\mu      &=& i \left( z_2 \partial_\mu z_1 -
                        z_1 \partial_\mu z_2
                 \right)
            = Z^T \sigma^y \partial_\mu Z     \; , \\
\bar B_\mu &=& i  \left( \bar z_1 \partial_\mu \bar z_2 -
                         \bar z_2 \partial_\mu \bar z_1
                  \right)
            = -\bar Z \sigma^y \partial_\mu \bar Z^T \; ,
\end{eqnarray}
which are related to the $SU(2)$ rotation via
$U^\dagger \partial_\mu U
= i \sigma^z A_\mu + i\bar B_\mu \sigma^+ + i B_\mu \sigma^-$.
Therefore we obtain after
applying the transformations (\ref{trafo}) to the action (\ref{labor})
the following alternative one:
\begin{eqnarray}
\label{local}
S_F&=&\int_0^\beta d\tau\int d^2r \:
\biggl\{
       p^\dagger \left( D_\tau^F + 2\tilde \gamma_\tau K_\tau \right) p
    +  \frac{1}{2m} \bar D_r^F  p^\dagger \; D_r^F p \nonumber \\
   &+& \tilde \gamma
       \left[2\left(\partial_r p^\dagger K_r p
                   - p^\dagger K_r \partial_r p
             \right)
             - 4 p^\dagger K_r K_r p
       \right]
\biggr\} \; ,
\\
\label{cp1}
S_{CP^1}&=&\frac{2}{g}\int_0^\beta d\tau\int d^2r \:
\left\{
       \bar D_r^B \bar Z \; D_r^B Z
       +\frac{1}{c^2}\bar D_\tau^B \bar Z \; D_\tau^B Z
\right\}
\end{eqnarray}
with $D_\mu^B=(\partial_\mu + i\; A_\mu )$ and $D_\mu^F=
(\partial_\mu + i\; \sigma^z A_\mu )$ denoting covariant derivatives
for the bosons and fermions, respectively, and $\tilde
\gamma_\tau = \gamma_\tau+1$ and
$\tilde \gamma= \gamma + 1/2m$. The action in the rotating reference frame is
invariant under a $U(1)$ transformation with $A_\mu$ being the corresponding
gauge connection \cite{wuze}. We would like to stress that the fermionic
fields in Eq.\ (\ref{local}) are the physical ones and not due to an
enlargement of the Hilbert space like in a slave-boson treatment of
the $t-J$ model \cite{naga}. The gauge fields arise by relating an $SU(2)$
rotation in spin-space and a vector on the sphere $S^2$. The manifold
$SU(2)$ is isomorphic to $S^3$, however, the vector $\vec{n}$ in $S^2$ fixes
only two of the three angles in $S^3$, and hence, a phase remains free.
\par
The dynamics of the gauge fields present in Eqs.~(\ref{local}) and
(\ref{cp1}) is
generated by fluctuations of the $CP^1$
non-linear $\sigma$ model as well as by the fermions
\cite{naga,dadda}, which can be systematically treated within a large-$N$
expansion \cite{dadda}. In contrast to earlier calculations on
related gauge theories \cite{wen}, where nondiagonal contributions like
the terms $\propto K_\mu$ in Eq.~(\ref{local}) were usually omitted,
it will be shown that exactly these terms are responsible for
the appearance of a doping induced quantum phase transition in this model.
\par
We proceed further by introducing $L$ copies of $p_\uparrow$ and
$p_\downarrow$ fermions and extend the $CP^1$ fields to $CP^{N-1}$ ones,
where now $\bar{Z} = (\bar{z}_1,...,\bar{z}_N)$ with $N=2L$.
After properly normalizing the action and
rescaling the fields the partition function of the field theory
given by Eqs.~(\ref{local}) and (\ref{cp1}) reads:
\begin{eqnarray}
\label{generating functional}
& & {\cal Z} [Q_\mu] =
\int{\cal D} Z {\cal D} \bar Z {\cal D} p {\cal D} p^\dagger  \:
\prod_{\vec r,\tau} \: \delta \left(\bar Z Z - \frac{2N}{g}\right)
\nonumber \\
& & \cdot \exp \left\{
-S_F - S_{CP^{N-1}} +\int_0^\beta d\tau \int d^2r
       \frac{g}{2N} Q_\mu A_\mu
\right\} \; ,
\end{eqnarray}
where we have introduced a source $Q_\mu$ for the  gauge field $A_\mu$,
which we use to separate the fermionic contribution
to the gauge fields from the one coming from the $CP^{N-1}$ model,
by substituting $(g/2N) A_\mu$ by a functional derivative with respect
to the source field $Q_\mu$.
Integrating out the fermions leads to
$S_F = -N\; {\rm Tr} \ln \left[ \Delta_F^0 + \Delta_F^A + \Delta_F^B
                         \right]$,
where we have divided the fermionic contribution into three terms, the
first one is the free part, the second one contains
the gauge field $A_\mu$ and the third one contains the
off-diagonal $B_\mu$ terms :
\begin{eqnarray}
\Delta_F^0 &=& \left(\partial_\tau+\frac{1}{2m}\partial_r^2\right)  \\
\Delta_F^A &=& i\sigma^z \frac{\delta}{\delta Q_\tau}
        + i \sigma^z \frac{1}{2m}
           \left[ \left( \partial_r \frac{\delta}{\delta Q_r}  \right)
                   + \frac{\delta}{\delta Q_r} \partial_r
           \right]
        - \frac{1}{2m} \: \frac{\delta^2}{{\delta Q_r}^2}  \\
\Delta_F^B &=& i \tilde \gamma_\tau \left(\frac{g}{2N}\right)
 2 \left(B_\tau\sigma^- + \bar B_\tau\sigma^+\right)
+ \tilde \gamma \left(\frac{g}{2N}\right)
   2 \left[\left(B_r \sigma^- + \bar B_r \sigma^+\right)
           \stackrel{\leftrightarrow}{\partial_r}
     \right]
\nonumber \\
&+& \left(\frac{g}{2N}\right)^2 \tilde \gamma
     4 B_r  \bar B_r \; .
\end{eqnarray}
This division turns out to be useful, since
because of the underlying $SU(2)$ structure
the $A_\mu$ and $B_\mu$ terms in $S_F$ do not mix and we
therefore can treat them independent form each other.
We first show that the $B_\mu$ terms modify the $CP^{N-1}$ model:
\begin{eqnarray}
\label{bterme1}
S_F^{B_\mu}&=& \int d^{3}\!p\: N \left(\frac{g}{2N}\right)^2
4 \bar B_r(p) B_s(-p) \nonumber \\
& &\biggl[ \tilde \gamma \delta_{rs} \int d^{3}\!q\: G_F(q)
+\frac{1}{2} \tilde \gamma^2
\int d^{3}\!q\:  G_F(q) (2q+p)_r (2q+p)_s G_F(q+p)
\biggr] \nonumber \\
&+&\int d^{3}\!p\:  N \left(\frac{g}{2N}\right)^2 4
\bar B_\tau (p) B_\tau (-p)
\frac{\tilde \gamma_\tau^2}{2} \int d^{3}\!q\: G_F(q) G_F(q+p)
\end{eqnarray}
where $G_F(q)= 1/(i\nu_n-\vec q^2/2m)$
is the fermionic Greens function and
we have introduced $\int d^{3}\!q\:$ as a shorthand notation for
$1/\beta \sum_{\nu_n} \int d^2q/(2\pi)^2 $.
Here $p=(\vec p,\omega_n)$ denotes the wave vector
together with the Matsubara frequency.
Evaluating the integrals in Eq.~(\ref{bterme1}) in the limit
$\vec p\rightarrow 0$ and $\omega/\vert\vec p\vert \rightarrow 0$ and
noticing that $(g/2N)\bar B_\mu B_\nu = {\bar D}_\mu^B \bar Z\:
D_\nu^B Z$
we obtain for $S_F^{B_\mu}$ together with the bosonic
contribution the following generalized $CP^{N-1}$ model:
\begin{eqnarray}
\label{bterme2}
S^{gen.}_{CP^{N-1}}=S_{CP^{N-1}}+ S_{F}^{B_\mu}
=\int_0^\beta d\tau \int d^2r \:
\biggl\{ f_r^\delta{\bar D}^B_r \bar Z\; D_r^B Z
+\frac{1}{c^2} f_\tau^\delta
{\bar D}^B_\tau \bar Z\; D_\tau^B Z
\biggr\},
\end{eqnarray}
where we have defined the doping dependent coupling constants
$f_r^\delta=1+2g\:\delta\left(\tilde \gamma-2m\:\tilde {\gamma}^2\right)
=1-2g\delta\left(1+2\gamma\:m\right)\gamma$ and
$f_\tau^\delta=1+2g c^2 \tilde \gamma_\tau^2
\frac{m}{4\pi}\theta (\delta)$, and
$\theta$ is the usual step-function.
As we see from Eq.~(\ref{bterme2}) the $B_\mu$ terms
lead to coupling constants that are now doping dependent.
Before we can integrate out the bosonic variables,
we have to decouple the quartic terms. This is achieved by introducing a
Hubbard Stratonovich field $\lambda_\mu$,
which turns out to be equivalent to the $U(1)$ gauge field since it couples
linearly to the source field $Q_\mu$:
\begin{eqnarray}
& &\exp \left\{
\frac{g}{2Nf_r^\delta}\int_0^\beta d\tau\int d^2r
\left[ -(\bar Z \partial_\mu Z)
(\bar Z \partial_\mu Z)+ i Q_\mu (\bar Z \partial_\mu Z)\right]
\right\} \nonumber \\
& & =\int {\cal D} \lambda_\mu \exp \left\{
\int_0^\beta d\tau\int d^2r
\left[ -\frac{2f_r^\delta}{g} \lambda_\mu \lambda_\mu +
\frac{1}{\sqrt{N}} \lambda_\mu 2i (\bar Z \partial_\mu Z) +
\frac{1}{\sqrt{N}} \lambda_\mu Q_\mu -\frac{g}{8Nf_r^\delta}
Q_\mu Q_\mu \right]
\right\}
\end{eqnarray}
Here we have rescaled the $Z$ bosons in such a way
that the constraint now reads $\bar Z Z=2Nf_r^\delta/g$.
We further include this constraint via a Lagrange-multiplier
field $\alpha (\vec r,\tau)$ into the action and introduce a mass term
$M^2 \bar Z Z$, which is arbitrary at this stage.
After these modifications the bosonic variables are integrated out and
together with the remaining contribution from the fermionic part we get:
\begin{eqnarray}
\label{bosonic part}
S=N\mbox{Tr ln}\Delta_B -N\mbox{Tr ln}\left( 1+\Delta_F^A/\Delta_F^0\right)
-\int_0^\beta d\tau \int d^2r\:
\frac{i2\sqrt{N}f_r^\delta}{g} \alpha(\vec r,\tau)
=: \sum_{\nu =1}^\infty N^{1-\nu/2} S^{(\nu)}\; ,
\end{eqnarray}
where
\begin{eqnarray}
\Delta_B&=& - D_\mu^B D_\mu^B + M^2-\frac{i\alpha}{\sqrt{N}} \; .
\end{eqnarray}
The covariant derivative is now given by
$D_\mu^B=\partial_\mu + i\lambda_\mu/\sqrt{N}$ and
furthermore the functional derivative with respect to the source $Q_\mu$
in $\Delta_F^A$ is substituted by $\lambda_\mu/\sqrt{N}$.
In the limit of large-$N$ the following two contributions are important:
\begin{eqnarray}
\label{s1}
S^{(1)}=
i{\tilde\alpha}\left[\frac{2f_r^\delta}{g}-\int d^{3}\!q\: G_B(q) \right]
\; ,
\end{eqnarray}
where $G_B(p)=1/(\vec p^2 + w_n^2+M^2)$ is the bosonic Greens function, and
\begin{eqnarray}
\label{s2}
S^{(2)}&=& \frac{1}{2}\int d^{3}\!p\:
        \biggl\{
           \lambda_\mu (p)\;
                 \left[ \Pi_{\mu\nu}^F (p) + \Pi_{\mu\nu}^B (p) \right] \;
           \lambda_\nu (-p)
           -\alpha (p)\; F(p)\;
            \alpha (-p)
        \biggr\} \; ,
\end{eqnarray}
where $F(p)=\int d^{3}\!q\: G_B(q)G_B(q+p)$
and where we have denoted the individual
contributions of the bosons and the fermions to the polarization tensor of
the gauge field $\lambda_\mu$ by $\Pi_{\mu\nu}^B (p)$
and $\Pi_{\mu\nu}^F (p)$, respectively.
At this point we want to emphasize that without taking into account
the nondiagonal $K_\mu$ terms of the fermionic action (\ref{fermlab}) there
would appear no doping dependent parameter as $f^\delta_r$ in
the expression for $S^{(1)}$.
\par
In the large-$N$ limit, the saddle-point condition $S^{(1)}=0$ has to be
imposed \cite{dadda}, as can be seen from Eq.\ (\ref{bosonic part}) such
that a definite relationship between the mass $M$ and the temperature
as well as the parameters of the model is established
(for $\beta \gg 1$):
\begin{eqnarray}
\label{mass}
M=\frac{2}{\beta}\mbox{arcsinh}\left[
\exp\left\{-4\pi\beta \left(\frac{f_r^\delta}{g}
-\frac{1}{g_c}\right)\right\}\right]\; ,
\end{eqnarray}
where we have defined the critical coupling constant
$g_c=8\pi/\Lambda$ ($\Lambda$ is an ultraviolet cutoff
$\sim a$, the lattice constant).
Eq.~(\ref{mass}) is the relationship obtained for the
inverse correlation length
of the non-linear $\sigma$ model \cite{chn,sach},
and the same discussion applies here.
As pointed out first by Chakravarty et al.~\cite{chn}, three different
phases appear depending on whether the coupling constant is smaller, larger,
or equal to the critical one.
\par
By introducing the renormalized spin stiffness
$\rho=1/g - 1/g_c$ at zero doping, where the system is
in the N\'eel ordered phase (at $T=0$) \cite{chn}, and setting the
expression in the exponential of Eq.~(\ref{mass}) equal to zero, we
obtain the following result for the critical doping:
\begin{eqnarray}
\label{critical doping}
\delta_c =\frac{\rho}{2\gamma(1+2\gamma\;m)} \; .
\end{eqnarray}
An estimation of the critical doping using the parameters $m=3/2$,
$\gamma=1/12$ and the value $\rho=7.9\cdot 10^{-3}$ \cite{para}
yields $\delta_c\approx 3.8\%$.
This is the first central result of this paper.
We have also calculated the spin-wave velocity and obtain a
reduction by doping to only about 80\% of its undoped value.
\par
The same result as Eq.~(\ref{mass}) can be obtained by integrating out
the fermions in Eq.~(\ref{fermlab}) similarly to previous calculations on
doped antiferromagnets \cite{alem}. However, in this case additional
information can be obtained.
As can be seen from Eq.~(\ref{s2})
a kinetic energy term for the gauge field was dynamically generated.
Evaluating the function $F(p)$ and
the polarization tensor of the bosons and fermions in the
limit of $\vec p \rightarrow 0$ and
$\omega /\vert \vec p \vert \rightarrow 0$, we get
choosing the Coulomb gauge $\vec \nabla \cdot \vec \lambda = 0$ a massive
propagator for both the lagrange multiplier field $\alpha$ and the
time component of the gauge field $\lambda_\tau$:
\begin{eqnarray}
D^\alpha &=& \frac{1}{8\pi M}\\
D^\lambda_{\tau\tau}&=&\frac{{\vec p}^2+\omega^2}{{\vec p}^2}
\frac{1}{\frac{1}{24\pi M}\left({\vec p}^2+\omega^2\right)+
\frac{m}{2\pi}}
\end{eqnarray}
Therefore fluctuations of the constraint field $\alpha$ as well as the
fluctuations of the time component of the gauge field $\lambda_\tau$
produce only short-range density-density interactions.
On the other hand for the propagator of the spatial part of the
gauge field $\lambda_i$ turns out to be massless:
\begin{eqnarray}
\label{propagator}
D_{ij}(\vec p , \omega ) = \frac{1}{2}
\left(\delta_{ij}-\frac{p_i p_j}{\vec p^2}\right) \frac{M}{\frac{1}{48 \pi}
{\vec p}^{2} - i \rho\frac{M}{m}\left(\frac{\omega}{p v_F}\right)}\; .
\end{eqnarray}
This is essentially the same propagator that was found by Nagaosa and
Lee \cite{naga}. The spatial components of the gauge field produce
a long-range current-current interaction between the fermions and the
$Z$-bosons. A central difference, however, is given by the fact that the
mass $M$ of the spin-excitations determines the strength of the propagator,
and hence the strength of the interaction mediated by it.
In fact according to the discussion following Eq.~(\ref{s1})
the mass $M$, is essentially the gap measured in neutron scattering
experiments \cite{barz}. Therefore, bound states are only possible in
the quantum disordered phase with a finite mass $M$.
In this regime the physical spectrum of the
theory contains only states with zero charge with respect to the gauge
field (singlet states) such as $Z$-$Z$, $p$-$Z$, and $p$-$p$ bound states.
The $Z$-$Z$ bound states correspond to spin-waves around the
antiferromagnetic wavevector, with a gap in the excitation
spectrum. The $p$-$Z$ bound states are spinless charged excitations.
Thus, this scenario gives an alternative way to charge-spin separation,
where the bare excitations are just spin-$\frac{1}{2}$ fermions but the
renormalized ones are spinless. It should be remarked here that in our case
charge-spin separation results from an interaction that leads to the
formation of a bound state and not by spontaneous breaking of gauge-invariance
as e.g.\ in Ref.\ \cite{naga}. Finally, the $p$-$p$ bound state is a
singlet with a charge $2e$ leading to pairing. Hence, charge-spin
separation and pairing are intimately connected in our case and result from
the same interaction.
\par
In summary, we presented a field-theoretic description of a microscopic
model for HTC that explicitly takes into account the influence of doping on
the transition to a quantum disordered phase and, moreover, reveals an
intimate relationship between the spin-gap, charge-spin separation, and
pairing.
\par
We acknowledge support by the Deutsche
For\-schungs\-ge\-mein\-schaft under Project
No.~Mu 820/5-2.


\begin{references}
\bibitem{note1} A general treatment will be given elsewhere. However,
               this situation is closely given by realistic parameters
               of the microscopic model \cite{para}.
\bibitem{para} For this estimation we use the following microscopic
               parameters of the spin-fermion model, for which the
               fermionic mass $m$ is isotropic
               \cite{prb}: $J_k=3/4$, $\bar t=-0.5$ and $\tilde t=-1/8$.
               All parameters are given in units of the
               hopping between the Cu and O
               orbitals $t_{pd}=1.3\mbox{eV}$ of the original
               3-band Hubbard model.
\end{references}
\end{document}